%% file: main.tex
\documentclass[journal=jacsat,manuscript=article]{achemso}

\usepackage{chemformula} 
\usepackage[T1]{fontenc} 
\usepackage{comment}
\usepackage{tabularx}
\usepackage{multirow}
\usepackage{makecell}
\usepackage{booktabs}
\usepackage[table]{xcolor}

\definecolor{lgray}{gray}{0.9}

\author{Margaret R. Martin}
\affiliation[Tufts University]
{Department of Computer Science, Tufts University, Medford, MA 02155, USA}
\author{Soha Hassoun}
\affiliation[Tufts University]
{Department of Computer Science, Tufts University, Medford, MA 02155, USA}
\alsoaffiliation[Tufts University]
{Department of Chemical and Biological Engineering, Tufts University, Medford, MA 02155, USA}
\email{soha.hassoun@tufts.edu}

\title[GIF]
  {General Intelligence-based Fragmentation (GIF): A framework for peak-labeled spectra simulation }

\abbreviations{IR,NMR,UV}

\begin{document}

\begin{tocentry}
\includegraphics[width=\linewidth]{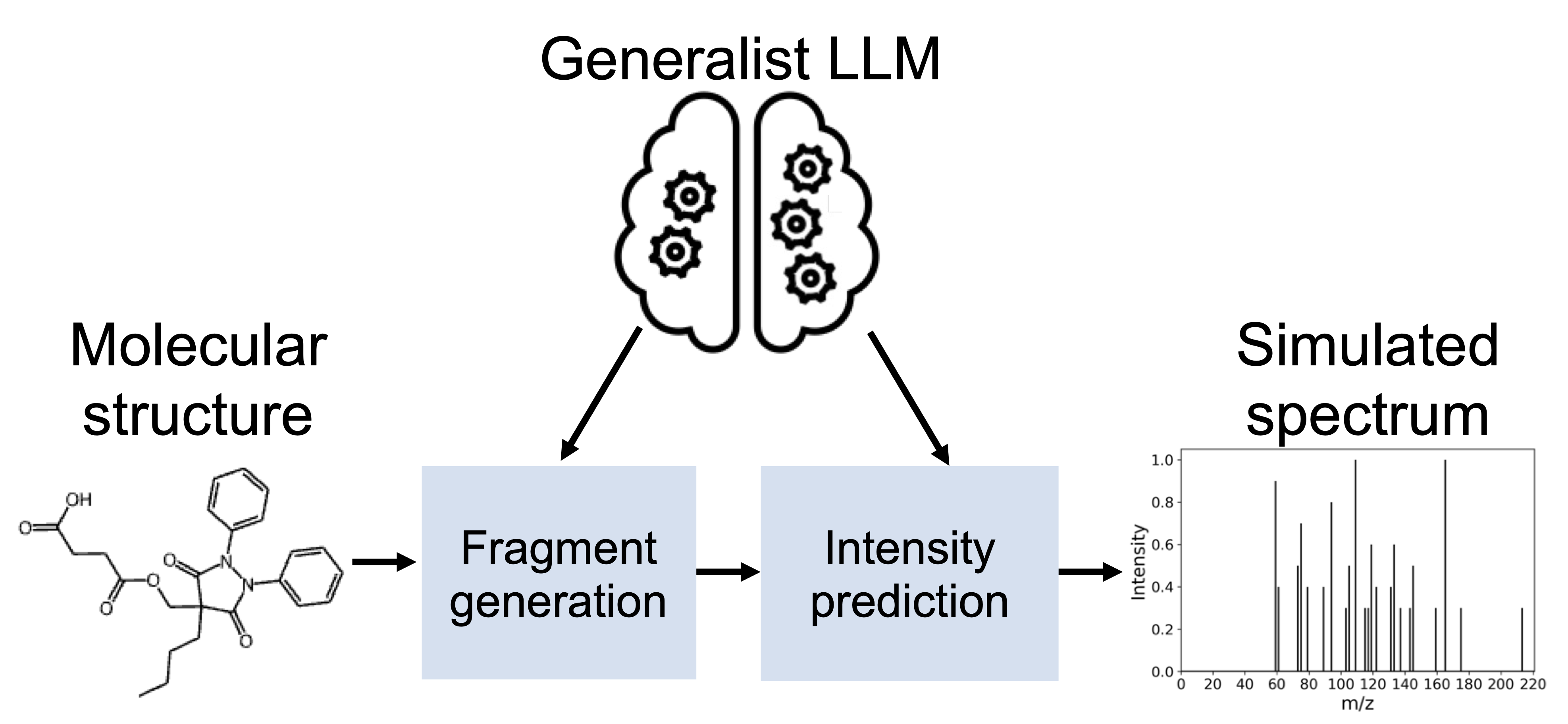}





\end{tocentry}

\begin{abstract}

Despite growing reference libraries and advanced computational tools, progress in the field of metabolomics remains constrained by low rates of annotating measured spectra.
The recent developments of large language models (LLMs) have led to strong performance across a wide range of generation and reasoning tasks, spurring increased interest in LLMs' application to domain-specific scientific challenges, such as mass spectra annotation.
Here, we present a novel framework, General Intelligence-based Fragmentation (GIF), that guides pretrained LLMs through spectra simulation using structured prompting and reasoning. GIF utilizes tagging, structured inputs/outputs, system prompts, instruction-based prompts, and iterative refinement. Indeed, GIF offers a structured alternative to ad hoc prompting, underscoring the need for systematic guidance of LLMs on complex scientific tasks. 
Using GIF, we evaluate current generalist LLMs' ability to use reasoning towards fragmentation and to perform intensity prediction after fine-tuning. We benchmark performance on a novel QA dataset, the MassSpecGym QA-sim dataset, that we derive from the MassSpecGym dataset.
Through these implementations of GIF, we find that GPT-4o and GPT-4o-mini achieve a cosine similarity of 0.36 and 0.35 between the simulated and true spectra, respectively, outperforming other pretrained models including GPT-5, Llama-3.1, and ChemDFM, despite GPT-5's recency and ChemDFM's domain specialization. GIF outperforms several deep learning baselines.
Our evaluation of GIF highlights the value of using LLMs not only for spectra simulation but for enabling human-in-the-loop workflows and structured, explainable reasoning in molecular fragmentation.
\end{abstract}

\section{Introduction}
Metabolomics plays a pivotal role in elucidating the chemical composition of biological samples, thereby enabling phenotyping, biomarker discovery, and research on health and disease. However, progress in metabolomics is often hindered by low spectra annotation rate, leaving many detected metabolites without chemical nor molecular identities. Traditionally, annotation involves matching experimental spectra to those in reference libraries such as GNPS\citep{wang2016sharing} and NIST23\citep{NIST23}. However, due to experimental costs and efforts, the size and coverage of reference libraries are limited. 
Further, as fragmentation patterns depend on experimental settings, e.g., collision energy, the same metabolite will produce differing spectra based on such settings. This variability further confounds annotation. 
To expand coverage, in silico approaches simulate spectra for putative candidate molecules that lack experimental references. Simulated spectra are then compared against the query spectrum to identify the most plausible match. Despite recent advances, generating accurate simulated spectra remains a significant challenge.

An increasing number of spectra simulation methods are structured to first identify molecular fragments corresponding to spectral peaks and then to assign intensity values (or scores) to these fragments. Two exceptions are MassFormer\citep{young2024tandem} and ESP\citep{li2024ensemble}, which learn to directly generate a binned spectrum from the query molecule. Decoupling fragment prediction and intensity prediction aligns more naturally with the physical process of tandem mass spectrometry, 
where fragmentation and detection are sequential and mechanistically distinct.
Indeed, the fragments themselves correspond  to specific molecular substructures resulting from bond cleavages, while fragment ion intensity reflects the relative stability and abundance of those fragments. 
Current spectra simulation methods vary in how they incorporate chemical knowledge into the fragmentation process.
CFM-ID\citep{wang2021cfm} uses explicit, human-curated fragmentation rules, e.g., single bond breaks (C–C, C–N, C–O), to combinatorially enumerate fragments. CFM-ID then assigns scores to these fragments using learned probabilistic models that estimate the likelihood of observing each fragment given its parent ion and corresponding fragmentation pathway. FraGNNet\citep{young2024fragnnet} implements an algorithm for plausible bond breaking to generate fragments and uses a graph neural network to learn the likelihood of fragmentation events. In contrast, ICEBERG\citep{goldman2024generating} and FIORA\citep{nowatzky2025fiora} learn to generate fragments using data-driven approaches, without explicit chemical rules. ICEBERG learns to generate fragmentation pathways as directed acyclic graphs from data and uses a set transformer to assign intensity values to the resulting fragments. FIORA iteratively generates fragments by breaking bonds stepwise and is trained end-to-end to predict fragment intensity. 

We propose in this paper a novel reasoning-based approach for spectra simulation. Reasoning tasks are often multi-step problems that involve sequential decision-making, compositional logic, and the integration of context to produce structured outputs, and cannot be easily solved using simple mechanisms such as data recall or pattern matching\citep{kojima2022large}. Large language models (LLMs) have shown capacity for reasoning and generalization
\citep{wei2022chain,kojima2022large, bran2023chemcrow}. These capabilities suggest that LLMs may be able to reason through the fragmentation process, rather than relying on chemical rules or learned fragment generation. Given this potential, we conjecture that LLMs are well-suited for fragment generation. 

General-purpose LLMs are trained on a large corpus of human language\citep{touvron2023llama, achiam2023gpt} and have reasoning abilities\citep{wei2022chain, kojima2022large}, making them applicable broadly across domains. When evaluated on chemistry related tasks, they exhibit understanding of chemistry-domain rules\citep{guo2024can, bhattacharya2024large}.
In parallel, there are now domain-adapted LLMs that are trained for general molecular understanding and generalize on related downstream tasks (BioT5\citep{pei2023biot5}, MolCA\citep{liu2023molca}, and ChemDFM\citep{zhao2025developing}). 
Further, specialized LLMs are trained for specific chemistry tasks, such as molecular property prediction (MolecularGPT\citep{liu2024moleculargpt} and ChemBERTa\citep{chithrananda2020chemberta}), and
drug discovery and molecular generation (DrugGPT\citep{li2023druggpt}, DrugLLM\citep{liu2024drugllm}, FragLlama\citep{shen2024fragllama}).
For mass spectra annotation, MolPuzzle\citep{guo2024can} creates three sub-tasks that address molecular structure elucidation and develops a QA dataset to evaluate general-purpose LLMs' performance against a human baseline. SpectraLLM\citep{su2025language} is trained using instruction-style prompts to perform spectrum-to-structure prediction.
However, no prior work utilizes LLMs for spectra simulation.

In this paper, we evaluate how general-purpose LLMs can be prompted and fine-tuned for spectra simulation.
To systematically evaluate this potential, we introduce a novel framework, General Intelligence-based Fragmentation (GIF). GIF consists of two steps that are executed sequentially through structured LLM queries. The fragmentation step leverages the LLM's reasoning abilities to identify likely fragments, while the intensity prediction step is achieved via fine-tuning the model to estimate relative fragment abundance. 
Each of these steps is performed iteratively, allowing the model to reflect and to perform iterative refinement\citep{elhenawy2024eyeballing, madaan2023self, iklassov2024self}. 
We develop a novel QA dataset derived from the MassSpecGym dataset\citep{bushuiev2024massspecgym}, and evaluate GIF using multiple pretrained LLMs and benchmark the results against previous methods. 
We present an example application of GIF where a user can query the LLM to generate fragments for candidate molecules and assess the similarity of the generated spectra against a query spectra. 
The contributions of this work are as follows: 
\begin{enumerate}
    \item Framing fragmentation as a reasoning task suited for general-purpose LLMs and introducing GIF, a fragment-then-score framework that incorporates structured prompting, fine-tuning, and iterative refinement to guide the application of general-purpose LLMs for spectra simulation. 

    \item Developing the MassSpecGym QA-sim dataset, a novel QA dataset to evaluate LLMs on spectra simulation through two structured molecule–spectrum tasks: fragment generation and intensity prediction.
  
    \item Implementing and evaluating GIF using multiple pretrained LLMs on the MassSpecGym QA-sim dataset. 
    GPT-4o and GPT-4o-mini achieve significantly higher performance compared to other general models, GPT-5, Llama-3.1, and to the chemistry-adapted ChemDFM.
    A single refinement step boosts fragment accuracy from below 1\% to 31.63\% and 24.37\% for GPT-4o and GPT-4o-mini, respectively, with continued improvement over multiple iterations. When combined with fine-tuning, cosine similarity increases from under 0.05 to 0.36 and 0.35, highlighting the effectiveness of targeted prompting and refinement. Importantly, we demonstrate that GIF achieves non-trivial performance in this difficult scientific simulation task when compared to other spectra simulation methods. 

    \item 
  Through substructure peak labeling, GIF transforms spectra annotation from an opaque prediction process into an interactive and explainable one. By enabling users to query, interpret, and refine fragment assignments, LLMs recast annotation as a collaborative reasoning task between human expertise and model inference. Human oversight can then be selectively applied to increase confidence and deepen interpretability when needed.
\end{enumerate}

\begin{figure*}[t]
\includegraphics[width=\linewidth]{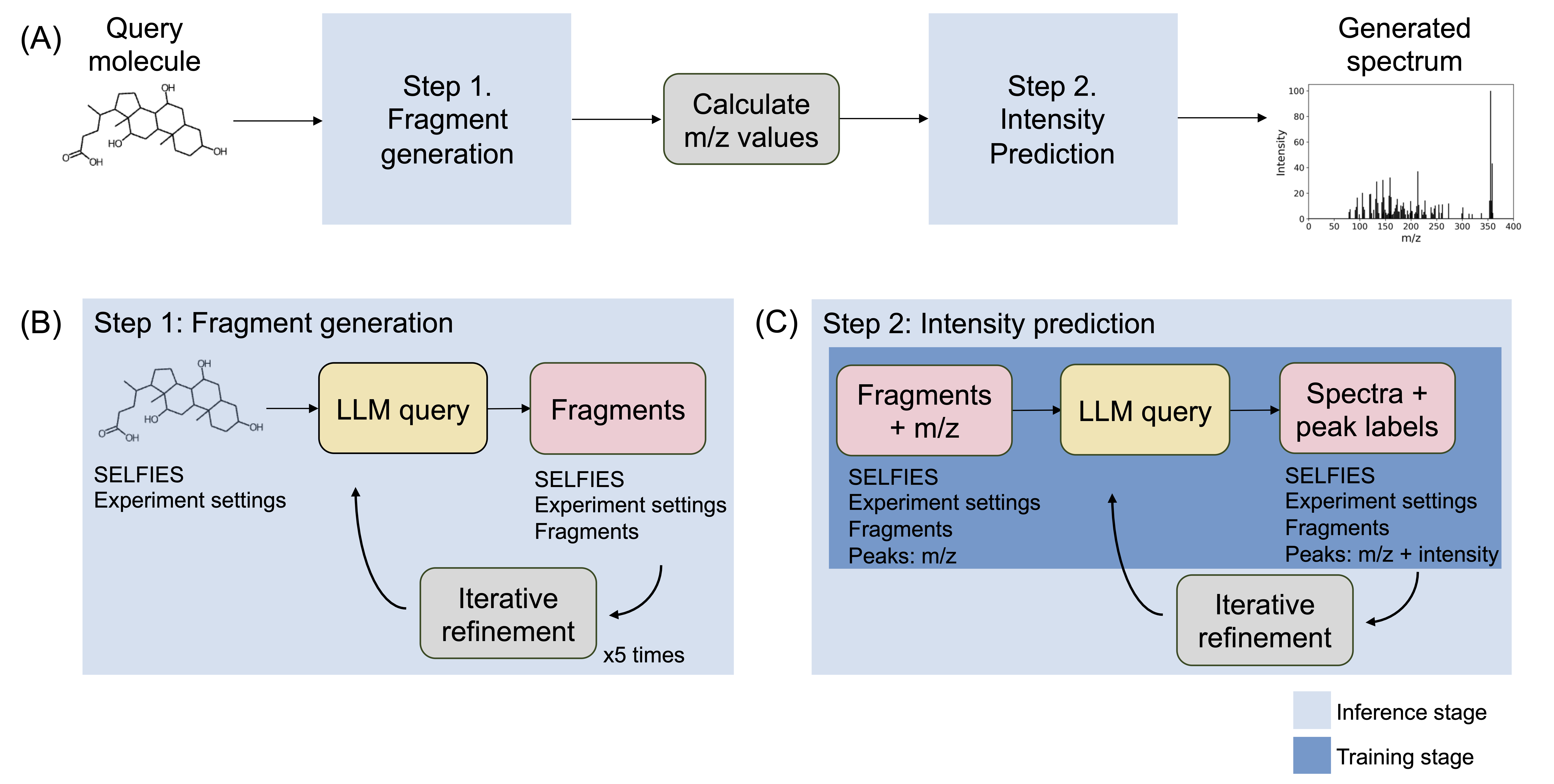}
\caption{GIF overview. Grey boxes represent offline processing steps, yellow boxes represent LLM queries, and pink boxes represent data. 
(A) GIF has two steps, fragment generation and intensity prediction, with an offline step for calculating the m/z values of the fragments.
(B) Fragment-generation step: 
Given the query molecule and the relevant experiment settings, the LLM suggests SELFIES fragments, which are filtered and used to prompt the LLM anew through iterative refinement.  
(C) Intensity-prediction step: Given the results of the first step, the LLM is prompted to predict the intensity values. The base LLM in this step is fine-tuned on intensity prediction and the prompting process utilizes a refinement step.}
\label{fig:method_pipeline}
\end{figure*}

\section{Methods}


\subsection{Method overview}
\label{sec:method_overview}
GIF consists of two steps (Figure \ref{fig:method_pipeline}A). 
The first step focuses on fragmentation 
(Figure \ref{fig:method_pipeline}B). 
Given a query molecule, GIF first generates a list of likely fragments, calculates the m/z values of the corresponding simulated mass spectrum peaks, and then predicts the corresponding intensity value of each peak. The predicted spectrum comprises the m/z values and predicted intensities form the simulated spectrum. Further, each generated peak has a fragment label reflecting the molecular fragment's structure, expressed as a SELFIES string. 
Fragment generation is performed by querying an LLM with the SELFIES representing the query molecule and the relevant experiment settings. The generated fragments are then filtered and used to create a new LLM query to refine the list of generated fragments. Filtering for invalid fragments and query refinement are performed five times to create the final list.  
The second step of GIF focuses on intensity prediction (Figure \ref{fig:method_pipeline}C), and utilizes an LLM that is fine-tuned for intensity prediction.
Given all collected information determined up to this point (SELFIES of query molecule, experiment settings, generated fragments, m/z values of simulated peaks), a fine-tuned LLM is prompted to predict the intensities. 
The LLM query response contains the predicted intensity values, and this response is cleaned and used to create a new LLM query to refine the peak intensity predictions. The refinement step is performed once to yield the finalized simulated spectrum with refined intensity values, completing the GIF process.

\subsection{Prompt design}
\label{sec:prompt_design}
To maximize LLM performance, we apply similar prompting strategies across both steps, including system prompts, instruction-based formatting, JSON-structured inputs and outputs, tagging, and temperature control, all techniques that have previously shown effective in prompting LLMs. 
We utilize system prompts to specify the domain of the query and a simplified summary of the task, fragmentation generation vs intensity prediction. System prompting yields consistent and higher-quality assistant responses\citep{djeffal2025reflexive, schulhoff2024prompt}. 
User prompts contain the specific instructions and the input data. The assistant response reflects the LLM's generated output. 


To support consistency and standardized analyses, the user's input data and the  assistant response are formatted as JSON objects.
In the user prompt, we use tags to specify the task (e.g., "\texttt{<\!<FragmentListPrediction>\!>}") and the provided data (e.g., "\texttt{<\!<MOL>\!>}" and "\texttt{<\!<EXP\_SETTINGS>\!>}"). Tagging increases performance when adapting general models to specialized domains, especially with fine-tuning\citep{shen2024tag}. 
Further, we format the prompts as instructions to guide the model's reasoning and perform instruction-tuning for increased task generalization\citep{shengyu2023instruction}.
Lastly, we vary the temperature of the user prompts. A high temperature (0.9) is used for the first four iterative refinement fragment generation steps to introduce more variety and novelty into the assistant response\citep{peeperkorn2024temperature}. All other prompts use a temperature of 0.1 so that the final response is generated more deterministically.
For molecular representations, the tokenization of SELFIES\citep{krenn2020self} is generally more robust towards generating chemically valid SELFIES as opposed to SMILES for general-purpose LLMs\citep{krenn2022selfies}. However, when evaluating GIF on ChemDFM\citep{zhao2025developing}, which is trained on SMILES\citep{weininger1988smiles}, we utilize SMILES instead of SELFIES. 



\subsection{Fragment generation}
Given a query molecule and the target experiment settings, the system and user prompts are iteratively queried to generate a list of likely fragments in the assistant response (Figure \ref{fig:method_pipeline}B). 
Here, the user prompt contains the SELFIES of the query molecule and the experiment settings formatted as a dictionary, containing the adduct, instrument, and collision energy. 
The desired assistant response is a list of fragments in order of descending intensity represented as SELFIES strings in JSON format. Each fragment in the output of this first query is filtered based on whether the fragment can be converted to a valid mol object using RDKit and whether that mol object represents a valid substructure of the query molecule. The list of valid fragments after filtering and the number of invalid fragments, as well as the SELFIES of the query molecule and the experiment settings again, are used to make a new user prompt that is queried with the corresponding system prompt to the base LLM to generate a new list of fragments.
This iterative prompting is done 5 times, and the output of the last query is filtered one last time and becomes the final list of fragments. An example of prompting and responses for fragment generation is provided as the first step in Figure \ref{fig:application}(A). 

\subsection{Intensity prediction}
The second GIF step is intensity prediction to determine the final simulated spectrum (Figure \ref{fig:method_pipeline}C). The intensities are discretized on a scale from 1 to 10 to improve performance.  The user prompt includes the SELFIES of the query molecule, the experiment settings, the generated fragments, and the calculated m/z values of the generated fragments.

As each fragment represents a peak in the simulated spectrum, the corresponding m/z value of each fragment is calculated offline using the mass of the fragment and the adduct. The generated fragments are listed in the same order as they were generated, as they were queried to be listed in order of descending intensity in step 1. The intensity values are discretized, and the prompt requests an intensity value for each m/z value on a scale of 1 to 10. The desired output of the LLM query is a list of dictionaries in JSON format where each entry contains the m/z and intensity values. 

The initial response of the LLM is post-processed and cleaned. Specifically, any peak entries are removed if the returned m/z value is not included in the query, the m/z value is a duplicate, or the intensity value is invalid.
The intensity values are merged with the previously queried list of generated fragments and corresponding m/z values.
A new user prompt is created which includes the new fragment list, the query molecule, and the experiment settings for one iterative refinement step.
The output of this query is the final output of GIF and represents the simulated spectrum. An example of prompting and responses for intensity prediction generation is provided as the second step in Figure \ref{fig:application}(A).

To enhance intensity prediction, we fine-tune GPT-4o-mini, GPT-4o, and Llama-3.1 in a supervised manner using the MassSpecGym QA-sim dataset. To explore performance, the models are trained using random subsets of the dataset of varying sizes. For GPT models, training is performed using API calls. For the Llama model, we use LLaMA-factory\citep{zheng2024llamafactory} to perform LoRA\citep{hu2022lora}. 
Because LLMs are trained for next-token prediction and perform best with predictable and informative sequences, we order the fragments by descending intensity during fine-tuning, where higher intensity fragments appear earlier in the sequence. This ordering matches that used in step 1 during generation. 
Further, we order the data in the training set by increasing complexity, which creates a specific type of curriculum learning scenario that has been shown to decrease runtime, improve training convergence, and increase performance\citep{kim2024strategic, li2022stability}. In our case, we calculate complexity as the sum of the normalized number of tokens in the prompt and the normalized number of fragments, where each value is divided by its respective maximum value.

\subsection{Pretrained LLMs}
We implement GIF using multiple state-of-the-art pretrained general-intelligence LLMs: GPT-4o-mini, GPT-4o, and GPT-5 from OpenAI\citep{achiam2023gpt} and Llama-3.1-8B-Instruct (Llama-3.1) from Meta\citep{grattafiori2024llama}. The architectures of these models are decoder-only transformers, except GPT-5, which is composed of a system of cooperating decoder-only sub-modules. Each such sub-model is an LLM trained in an autoregressive manner for next-token prediction but vary in size and specialization. A real-time router dynamically determines which sub-models are activated for a given input.
We also evaluate ChemDFM\citep{zhao2025developing}, which uses a pretrained Llama-13B as a base model with additional chemistry-domain pretraining on literature and instruction tuning on molecular and biochemical tasks. As a result, ChemDFM is similar in architecture, intelligence, and reasoning to the other included models, but has additional domain knowledge. 
The relevant code and detailed instructions are available as open source under the MIT license at https://github.com/HassounLab/GIF.

\section{Results and discussion}

\subsection{MassSpecGym QA-sim dataset}
To train and evaluate GIF, we develop a novel QA dataset from the MassSpecGym benchmark dataset\citep{bushuiev2024massspecgym} that we refer to as the MassSpecGym QA-sim dataset. This dataset is designed to support the reasoning then scoring paradigm utilized by GIF. For each unique spectrum in the MassSpecGym dataset, we create a QA pair for fragment generation and another pair for intensity prediction. We apply the designed prompts as templates to create the QA pairs from MassSpecGym. The only additional data required that are not present in MassSpecGym are the substructure peak labels, which we calculate using MAGMa\citep{ridder2014automatic}. From a query molecule, MAGMa combinatorially breaks bonds and applies neutral losses to generate fragments, matches those fragments to peaks in the query spectrum, and then assigns scores to the matches. 
The resulting MassSpecGym QA-sim dataset consists of 462,208 QA pairs, derived from the 231,104 annotated spectra in MassSpecGym. 
Following the specified training split for MassSpecGym, 213,548 and 17,556 of the QA pairs for each task are assigned to training and testing sets, respectively. In this work, we utilize random subsets of the training set of varying sizes to fine-tune for intensity prediction, and we evaluate GIF and report results on the test set for fragment generation and intensity prediction. 
As these QA pairs have been developed for LLMs and fine-tuning, they can be utilized as a benchmark dataset covering the MassSpecGym dataset for the development of future LLM models for annotation.

\subsection{Evaluation of pretrained LLMs}
We first assess the performance of several pretrained LLMs on GIF. For this evaluation, we simulate spectra for a random subset of 300 query molecules in the test set of MassSpecGym using GPT-4o-mini, GPT-4o, GPT-5, Llama-3.1, and ChemDFM. We select the GPT models to evaluate small and large sizes of generalist models. We  include Llama-3.1 as a generalist model sized between GPT-4o-mini and GPT-4o. We include ChemDFM as a domain-specific model. We evaluate the results using three metrics. Subformula accuracy captures the proportion of "true" subformulae, as determined by MAGMa, of the generated fragments. It is calculated with respect to the total number of fragments determined by MAGMa. The cosine similarity and Jensen-Shannon similarity capture
angular and distributional similarity, respectively, and are calculated as specified in MassSpecGym\citep{bushuiev2024massspecgym}. Both metrics compare the similarity between two vectorized spectra and are on a scale from 0 to 1, where 1 represents a perfect match. 

Across all metrics, GPT-4o and GPT-4o-mini achieve the highest and second highest results, respectively (Table \ref{tab:each_llm_best_performance}). When using the other models, the subformula accuracy is below 5\%, naturally leading to lower cosine and Jensen-Shannon similarities. As fine-tuning is not currently available for GPT-5, it is not evaluated on the intensity prediction step. The low scores on subformula accuracy suggests potential issues in routing the query to an appropriate expert sub model within GPT-5. 
ChemDFM, the only in-domain model, resulted in the lowest performance, despite pretraining on the Llama-13B model. This finding suggests that reasoning and task generalization  abilities of the GPT models is more important for this application than the chemistry knowledge. 

\begin{table}[!ht]
\scriptsize
\setlength{\tabcolsep}{4pt} 
    \centering
    \begin{tabular}{l c c c }
    \toprule
    Pretrained LLM &  Subformula accuracy (\%) & Cosine similarity & Jensen-Shannon similarity \\
    \midrule
     GPT-4o-mini &39.05 &0.35 &0.38 \\
     GPT-4o &61.60 &0.36 &0.37 \\
     GPT-5 &2.58 &\cellcolor{lgray} &\cellcolor{lgray} \\
     Llama-3.1 &4.62 &0.01 &0.01 \\
     ChemDFM & 0.99 &0.00 &0.00 \\
 
    \bottomrule
    \end{tabular}

\vspace*{5mm}
\caption{Comparison of GIF using several pretrained LLMs using subformula accuracy, cosine similarity, and Jensen-Shannon similarity. This evaluation is on a random subset of 300 query molecules in MassSpecGym. }
\label{tab:each_llm_best_performance}
\end{table}

\subsection{Ablation studies}
To evaluate the impact of iterative refinement and fine-tuning, we perform ablation studies (Table \ref{tab:ablation_study}) on the same random subset of data as the previous analysis. As GPT-4o and GPT-4o-mini are the only base models for GIF that achieve a cosine similarity above 0.1, we assess these models for the ablation studies. We report on additional metrics. The fragment accuracy represents the proportion of "true" fragments that were generated. It is calculated as the number of generated fragments that exhibit the same two-dimensional structure of the MAGMa fragments divided by the total number of MAGMa fragments. Chemical validity represents the proportion of the generated SELFIES that are valid strings. We confirmed validity by converting SELFIES to SMILES using the SELFIES package\citep{lo2023recent} and then to a valid Mol object using RDKit\citep{landrum2013rdkit}. Lastly, substructure validity is the proportion of generated fragments that are true substructures of the query molecule, determined using RDKit. Fragment accuracy, subformula accuracy, chemical validity, and substructure validity act as metrics for evaluating fragment generation results of step 1, whereas cosine similarity and Jensen-Shannon similarity act as metrics evaluating intensity prediction results of step 2.

We evaluate the impact of iterative refinement for fragment generation by reporting the relevant metrics for a baseline model and after each of the five iterative refinement steps. 
Using GPT-4o and GPT-4o-mini, GIF using the base model without iterative refinement or fine-tuning achieves low results. For example, the fragment accuracy is less than 1\% for GPT-4o-mini and GPT-4o, and one step of iterative refinement increases the fragment accuracy to 24.37\% and 31.63\%, respectively. For each step of iterative refinement for fragment generation, the metrics continue to improve. The only exceptions are chemical validity and substructure validity, but they remain  above 98\% for subsequent iterative refinement steps. 

For intensity prediction, we evaluate the impact of fine-tuning using a select number of data points, 500, 1000, 2000, 5000, 10000, where the larger sets were needed for fine-tuning the smaller models. We also evaluate the impact of including iterative refinement. 
Whether only iterative refinement or only fine-tuning is applied, cosine similarity is no higher than 0.06. However,  when both are applied, cosine similarity is at least 0.18. Examining the results without fine-tuning, the model neglects to include the exact queried m/z values in the response.  We limited iterative refinement to one step as additional steps lowered the results. 
The ablation studies also show the effect of the number of training data points on the result. Our results suggest that fine-tuning with too many training points leads to overfitting. The best performing fine-tuned model was trained on 5,000 points for GPT-4o-mini and on 1,000 points for GPT-4o.

\begin{table}[!ht]
\scriptsize
\setlength{\tabcolsep}{4pt} 
    \centering
    \begin{tabular}{l c c c c c c c }
    \toprule
    Method & \makecell{Fragment\\accuracy(\%)} & \makecell{Subformula\\accuracy(\%)} & \makecell{Chemical\\validity(\%)} & \makecell{Substructure\\validity(\%)} & \makecell{Cosine\\similarity} & \makecell{Jensen-Shannon\\similarity} \\
    \midrule
    \multicolumn{7}{c}{(A) GPT-4o-mini} \\ \midrule

    Base model & 0.03 & 9.56 & 72.33 &45.33 &0.04 &0.03 \\
    Base model + 1 FragIR & 24.37 & 32.75 & 93.00 & 84.67 &\cellcolor{lgray} &\cellcolor{lgray} \\
    Base model + 2 FragIR & 28.38 & 35.67 & 97.67 & 94.33 &\cellcolor{lgray} &\cellcolor{lgray} \\
    Base model + 3 FragIR & 31.26 & 37.84 & 99.00 & 93.33 &\cellcolor{lgray} &\cellcolor{lgray} \\
    Base model + 4 FragIR & 31.65 & 38.45 & 99.33 & 99.00 &\cellcolor{lgray} &\cellcolor{lgray} \\
    Base model + 5 FragIR & \textbf{32.34} & \textbf{39.05} & \textbf{99.67} & \textbf{99.33} &  0.05 & 0.05 \\

    Base model + 5 FragIR + 1 IntIR &\cellcolor{lgray} &\cellcolor{lgray} &\cellcolor{lgray} &\cellcolor{lgray} & 0.21 & 0.21 \\
    Base model + 5 FragIR + FT 1000 &\cellcolor{lgray} &\cellcolor{lgray} &\cellcolor{lgray} &\cellcolor{lgray} & 0.06 & 0.06 \\
    Base model + 5 FragIR + FT 1000 + 1 IntIR &\cellcolor{lgray} &\cellcolor{lgray} &\cellcolor{lgray} &\cellcolor{lgray} & 0.18 & 0.18 \\
    Base model + 5 FragIR + FT 5000 &\cellcolor{lgray} & \cellcolor{lgray}&\cellcolor{lgray} &\cellcolor{lgray} & 0.05 & 0.07 \\
    Base model + 5 FragIR + FT 5000 + 1 IntIR &\cellcolor{lgray} &\cellcolor{lgray} &\cellcolor{lgray} &\cellcolor{lgray} & \textbf{0.35} & \textbf{0.38} \\
    Base model + 5 FragIR + FT 10000 &\cellcolor{lgray} &\cellcolor{lgray} &\cellcolor{lgray} &\cellcolor{lgray} & 0.06 & 0.12 \\
    Base model + 5 FragIR + FT 10000 + 1 IntIR &\cellcolor{lgray} &\cellcolor{lgray} &\cellcolor{lgray} &\cellcolor{lgray} & 0.20 & 0.22 \\

    \midrule
    \multicolumn{7}{c}{(B) GPT-4o} \\ \midrule

    Base model & 0.17 &5.81 &97.00 &55.00 &0.01 &0.01 \\ 
    Base model + 1 FragIR & 31.63 & 34.74 &98.67 &72.00 &\cellcolor{lgray} &\cellcolor{lgray} \\
    Base model + 2 FragIR & 44.93 &47.11 &\textbf{100.00} &89.67 &\cellcolor{lgray} &\cellcolor{lgray} \\
    Base model + 3 FragIR & 55.21 &57.65 &99.67 &93.00 &\cellcolor{lgray} &\cellcolor{lgray} \\
    Base model + 4 FragIR & 57.67 &60.06 &99.67 &\textbf{99.33} &\cellcolor{lgray} &\cellcolor{lgray} \\
    Base model + 5 FragIR & \textbf{59.66} &\textbf{61.60} &99.67 &98.33 &0.04 &0.04 \\

    Base model + 5 FragIR + 1 IntIR &\cellcolor{lgray} &\cellcolor{lgray} &\cellcolor{lgray} &\cellcolor{lgray} & 0.04 &0.05 \\
    Base model + 5 FragIR + FT 500 &\cellcolor{lgray} &\cellcolor{lgray} &\cellcolor{lgray} &\cellcolor{lgray} & 0.05 & 0.06 \\
    Base model + 5 FragIR + FT 500 + 1 IntIR &\cellcolor{lgray} &\cellcolor{lgray} &\cellcolor{lgray} &\cellcolor{lgray} & 0.24 &0.25 \\
    Base model + 5 FragIR + FT 1000 &\cellcolor{lgray} & \cellcolor{lgray}&\cellcolor{lgray} &\cellcolor{lgray} &  0.05 &0.06 \\
    Base model + 5 FragIR + FT 1000 + 1 IntIR &\cellcolor{lgray} &\cellcolor{lgray} &\cellcolor{lgray} &\cellcolor{lgray} &  \textbf{0.36} &\textbf{0.37} \\
    Base model + 5 FragIR + FT 2000 &\cellcolor{lgray} &\cellcolor{lgray} &\cellcolor{lgray} &\cellcolor{lgray} & 0.05 &0.06 \\
    Base model + 5 FragIR + FT 2000 + 1 IntIR &\cellcolor{lgray} &\cellcolor{lgray} &\cellcolor{lgray} &\cellcolor{lgray} & 0.18 & 0.20 \\

    \bottomrule
    \end{tabular}

\vspace*{5mm}
\caption{Ablation studies to evaluate the impact of iterative refinement and fine-tuning for GIF when using (A) GPT-4o-mini and (B) GPT-4o. The number of iterative refinement steps used for fragment generation and intensity prediction are denoted by "FragIR" and "IntIR", respectively. If fine-tuning is applied, it is denoted by "FT" and the following number is the number of training and validation data points used. As iterative refinement of fragment generation directly affects the results of step 1, we report the corresponding first four metrics only. Similarly, we report the final two metrics when applying fine-tuning and iterative refinement of intensity prediction.
The best performance of each metric for each section of the table are indicated in bold. }
\label{tab:ablation_study}
\end{table}

\begin{table}[!ht]
\scriptsize
\setlength{\tabcolsep}{4pt} 
    \centering
    \begin{tabular}{l c c  }
    \toprule
    Method & Cosine similarity & Jensen-Shannon similarity \\
    \midrule
    Precursor m/z & 0.15 &0.59 \\
    FFN Fingerprint &0.25 &0.69 \\
    GNN &0.19 &0.64 \\
    FraGNNet &0.52 &0.91 \\
    GIF &0.35 &0.37 \\

    \bottomrule
    \end{tabular}

\vspace*{5mm}
\caption{Benchmark performance of GIF against baseline methods evaluated on the MassSpecGym test set.}
\label{tab:benchmark_performance}
\end{table}

\subsection{Benchmark results}
GPT-4o-mini and GPT-4o achieved comparable cosine and Jensen-Shannon similarity scores. Due to cost considerations,  we implement GIF using GPT-4o-mini rather than GPT-4o to simulate spectra for all 17,556 test data points in MassSpecGym. Our results (Table \ref{tab:benchmark_performance}) are compared to previous baseline methods as reported in the MassSpecGym publication\citep{bushuiev2024massspecgym}.
Precursor m/z is a simplified method that simulates a single-peak spectrum where the only peak represents the precursor m/z, which is calculated using the mass of the query molecule and the adduct. The other three methods are deep-learning approaches. FFN Fingerprint is a feedforward network that simulates the spectra based on the fingerprint representations of the query molecules. 
GNN is a Graph Isomorphism Network variant of a graph neural network, where the query molecule is represented as a 2D graph. FraGNNet\citep{young2024fragnnet} consists of first performing combinatorial fragmentation and then uses a GNN to learn a probability distribution over molecule fragments. Notably, GIF achieves higher cosine similarity than FFN Fingerprint and GNN. GPT-4o-mini guided by GIF therefore  outperforms deep learning baseline model. However, GIF does not outperform the current state-of-the-art method, FraGNNet, which achieves a cosine similarity of 0.52.

\subsection{Example application of GIF}
\begin{figure*}[t]
\includegraphics[width=\linewidth]{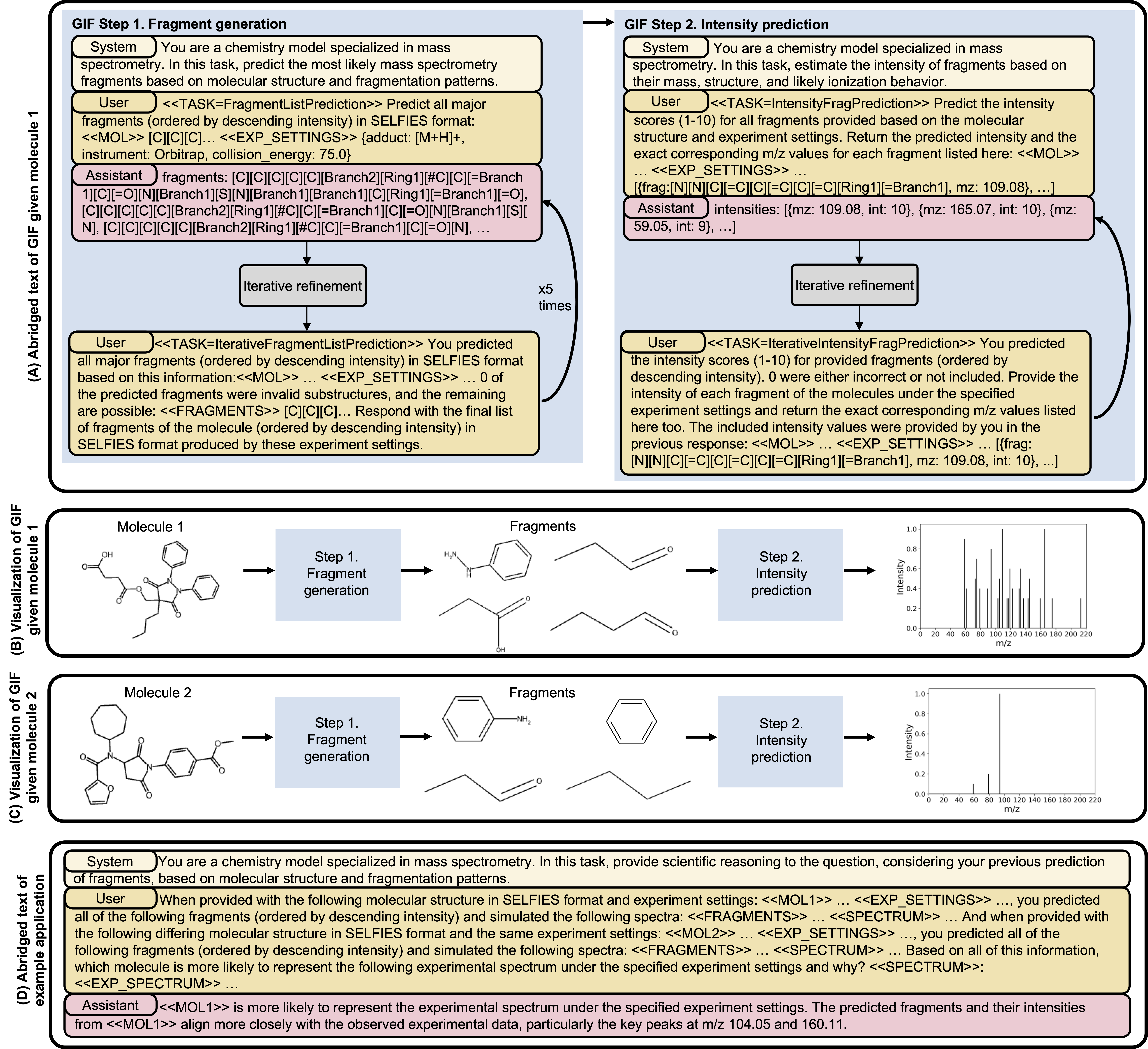}
\caption{An example of GIF guiding LLMs in spectra simulation of two candidate molecules. (A) An abridged textual representation of the application of GIF to molecule 1 (Suxibuzone) through steps 1 and 2. (B) The corresponding visualization of GIF's application to molecule 1. (C) The visualization of GIF's application to molecule 2 (Methyl 4-[3-[cycloheptyl(furan-2-carbonyl)amino]-2,5-dioxopyrrolidin-1-yl]benzoate). (D) An abridged textual representation of the example downstream application that uses the output of GIF when querying molecule 1 and molecule 2. 
The blue boxes represent GIF steps, either fragment generation or intensity prediction. The yellow boxes are prompts queried to the LLM, where the lighter box is the system prompt and the darker boxes are user prompts. The pink boxes are the LLM responses. We use GPT-4o for all queries in this example.
}
\label{fig:application}
\end{figure*}

To demonstrate the utility and interpretability enabled by GIF, we highlight its use in a real-world analysis scenario requiring spectrum–molecule reasoning and explainability (Figure \ref{fig:application}). A user is pondering which of two candidate molecules best matches a query spectrum. The user utilizes an LLM guided by GIF to simulate a spectrum for each candidate and then compares the simulated spectra to the query spectrum. Here, molecule 1 is the target molecule, Suxibuzone, and molecule 2 is a likely candidate molecule, Methyl 4-[3-[cycloheptyl(furan-2-carbonyl)amino]-2,5-dioxopyrrolidin-1-yl]benzoate.

We showcase the textual format used in each step of GIF as well as an example of the format of the assistant's response (Figure \ref{fig:application}A). In step 1, the system and user prompt are queried and a list of fragments is generated in the assistant response. The user prompt contains the data: molecular structure of molecule 1 and the experiment settings. Offline iterative refinement processing determines that 0 of the generated fragments were invalid substructures, and a new user prompt is queried with this information. The assistant response is a new list of generated fragments. Iterative refinement is performed 5 times. Step 2 assigns intensity values to generated fragments. The user prompt contains all necessary data, including molecular structure, experiment settings, and the generated fragments with corresponding m/z values. Offline iterative refinement determines that all of the fragments were labeled with a valid intensity value, and an additional user prompt is queried. GIF is applied separately to molecules 1 and 2 and varying results are generated (Figure \ref{fig:application}B and C). The user then creates a prompt that includes the GIF results and the experimental spectrum to query the LLM to determine which molecule is more likely to be represented in the experimental spectrum (Figure \ref{fig:application}D). The LLM reasons through the prompt and identifies the correct molecule. This last step leverages the LLM's reasoning abilities. It is formatted using our prompting strategy, and it is zero-shot, without prior training or fine-tuning of the LLM.

\subsection{Discussion}
While LLMs exhibit strong reasoning capabilities, applying them directly to spectra simulation introduces domain-specific challenges, such as representing molecular structures, incorporating experimental conditions, and generating chemically valid fragments. Developing appropriate
prompting strategies is challenging to the user, and may lead to inconsistent and poor outcomes.
The GIF framework addresses these issues by translating the task into a structured reasoning problem: it guides the LLM through prompting, validity checks, formatting to ensure consistency, and leverages iterative self-refinement to improve accuracy. To evaluate the capabilities of LLMs in mass spectra simulation and support reproducibility, we developed the MassSpecGym QA-sim dataset, a structured QA benchmark designed to test fragment generation and intensity prediction. 

The GIF framework is model agnostic as we showed that it can be implemented and evaluated using multiple pretrained LLMs. GIF achieves a cosine similarity of 0.35 when evaluated on the MassSpecGym test set using GPT-4o-mini. GPT-4o and GPT-4o-mini achieve the highest accuracy and cosine similarity, outperforming other pretrained models including GPT-5, Llama-3.1, and ChemDFM. Despite size and domain knowledge, using ChemDFM to deploy GIF results in the lowest cosine similarity of 0.00 and annotating the peaks with substructure labels at 0.99\% accuracy. These results suggest that reasoning abilities, and not domain-specific knowledge, have the greatest impact on performance.
When implemented with GPT-4o-mini, GIF is competitive with baseline deep learning methods, but does not outperform the current state-of-the-art method. 

Our ablation studies using GPT-4o and GPT-4o-mini underscore the importance of fine-tuning and iterative refinement in achieving high performance on spectra simulation. The first iterative refinement step of fragment generation increases the fragment accuracy from less than 1\% to 31.63\% and 24.37\% for GPT-4o and GPT-4o-mini, respectively. The combination of iterative refinement and fine-tuning is critical to enhance performance on intensity prediction, achieving a cosine similarity of 0.36 and 0.35 for GPT-4o and GPT-4o-mini, respectively. 

Beyond performance, GIF's implementation using LLMs provides an interpretable,  user-centered paradigm for spectra simulation. 
In our example application of GIF, we demonstrate how a user may include the GIF-based simulated spectra and substructures to annotate an experimental spectrum given two candidates.
GIF in this scenario exemplifies the practical application of models that exhibit understanding of language, molecule, and spectra. 
Further, the two-step structure of GIF supports "human-in-the-loop" interaction, enabling users to add or remove fragments before intensity prediction and final spectra simulation. This capability facilitates including additional domain knowledge and stopping model errors from continuing to downstream steps. Indeed, language-capable models play a crucial role in enhancing usability. The ability to follow instructions, support structured prompts, and generalize across tasks makes such models easy to deploy and adapt in real-world settings. These traits are vital for making computational tools more accessible, transparent, and collaborative in domains such as metabolomics.


To further advance the GIF framework, future work will prioritize improving performance and usability. Using reinforcement learning from human feedback (RLHF) instead of supervised fine-tuning in cases like fragmentation, where data is limited or qualitative, would likely achieve higher performance\citep{mandal2025distributionally, kaufmann2024survey}. Similarly, dynamic fine-tuning, which rescales the objective function based on token probability, has been shown to achieve higher generalization to unseen data than traditional supervised fine-tuning, and as a result, may increase the performance on intensity prediction\citep{wu2025generalization}. 
Tool integration with LLMs\citep{gao2025democratizing} can significantly streamline complex tasks such as mass spectra annotation. In the current GIF implementation, we calculated m/z values for the fragments offline 
in between queries as the LLMs inconsistently calculated such values. Structure validity checks were also performed offline. Structural annotation from MAGMa can be presented to the LLM as multiple-choice queries thus combining traditional tools with LLMs for annotation. Indeed, multiple-choice answers conceptually change the task from generation to classification and often leads to higher performance\citep{ghosal2022two}. Looking forward, we suggest that fine-tuning general-purpose LLMs with additional molecular and spectral data can enable deeper reasoning about fragmentation and spectra annotation.

\begin{acknowledgement}

Research reported in this publication was supported by the National
Institute of General Medical Sciences of the National Institutes of Health
under award number R35GM148219.
The content is solely the responsibility of the authors and does not necessarily represent the official views of the NIH.

\end{acknowledgement}

\input{main.bbl}

\end{document}

%% file: main.bbl
\providecommand{\latin}[1]{#1}
\makeatletter
\providecommand{\doi}
  {\begingroup\let\do\@makeother\dospecials
  \catcode`\{=1 \catcode`\}=2 \doi@aux}
\providecommand{\doi@aux}[1]{\endgroup\texttt{#1}}
\makeatother
\providecommand*\mcitethebibliography{\thebibliography}
\csname @ifundefined\endcsname{endmcitethebibliography}  {\let\endmcitethebibliography\endthebibliography}{}

%% file: main.bbl
\begin{mcitethebibliography}{50}
\providecommand*\natexlab[1]{#1}
\providecommand*\mciteSetBstSublistMode[1]{}
\providecommand*\mciteSetBstMaxWidthForm[2]{}
\providecommand*\mciteBstWouldAddEndPuncttrue
  {\def\EndOfBibitem{\unskip.}}
\providecommand*\mciteBstWouldAddEndPunctfalse
  {\let\EndOfBibitem\relax}
\providecommand*\mciteSetBstMidEndSepPunct[3]{}
\providecommand*\mciteSetBstSublistLabelBeginEnd[3]{}
\providecommand*\EndOfBibitem{}
\mciteSetBstSublistMode{f}
\mciteSetBstMaxWidthForm{subitem}{(\alph{mcitesubitemcount})}
\mciteSetBstSublistLabelBeginEnd
  {\mcitemaxwidthsubitemform\space}
  {\relax}
  {\relax}

\bibitem[Wang \latin{et~al.}(2016)Wang, Carver, Phelan, Sanchez, Garg, Peng, Nguyen, Watrous, Kapono, Luzzatto-Knaan, Porto, Bouslimani, Melnik, Meehan, Liu, Crüsemann, Boudreau, Esquenazi, Sandoval-Calderón, Kersten, Pace, Quinn, Duncan, Hsu, Floros, Gavilan, Kleigrewe, Northen, Dutton, Parrot, Carlson, Aigle, Michelsen, Jelsbak, Sohlenkamp, Pevzner, Edlund, McLean, Piel, Murphy, Gerwick, Liaw, Yang, Humpf, Maansson, Keyzers, Sims, Johnson, Sidebottom, Sedio, Klitgaard, Larson, and Boya~P]{wang2016sharing}
Wang,~M. \latin{et~al.}  Sharing and community curation of mass spectrometry data with Global Natural Products Social Molecular Networking. \emph{Nature biotechnology} \textbf{2016}, \emph{34}, 828--837\relax
\mciteBstWouldAddEndPuncttrue
\mciteSetBstMidEndSepPunct{\mcitedefaultmidpunct}
{\mcitedefaultendpunct}{\mcitedefaultseppunct}\relax
\EndOfBibitem
\bibitem[NIST(2023)]{NIST23}
NIST NIST Tandem Mass Spectral Library 2023 Release. 2023; \url{https://chemdata.nist.gov/dokuwiki/lib/exe/fetch.php?media=chemdata:asms2023:asms2023_nist23_features.pdf}\relax
\mciteBstWouldAddEndPuncttrue
\mciteSetBstMidEndSepPunct{\mcitedefaultmidpunct}
{\mcitedefaultendpunct}{\mcitedefaultseppunct}\relax
\EndOfBibitem
\bibitem[Young \latin{et~al.}(2024)Young, R{\"o}st, and Wang]{young2024tandem}
Young,~A.; R{\"o}st,~H.; Wang,~B. Tandem mass spectrum prediction for small molecules using graph transformers. \emph{Nature Machine Intelligence} \textbf{2024}, \emph{6}, 404--416\relax
\mciteBstWouldAddEndPuncttrue
\mciteSetBstMidEndSepPunct{\mcitedefaultmidpunct}
{\mcitedefaultendpunct}{\mcitedefaultseppunct}\relax
\EndOfBibitem
\bibitem[Li \latin{et~al.}(2024)Li, Zhou~Chen, Kalia, Zhu, Liu, and Hassoun]{li2024ensemble}
Li,~X.; Zhou~Chen,~Y.; Kalia,~A.; Zhu,~H.; Liu,~L.-p.; Hassoun,~S. An Ensemble Spectral Prediction (ESP) model for metabolite annotation. \emph{Bioinformatics} \textbf{2024}, \emph{40}, btae490\relax
\mciteBstWouldAddEndPuncttrue
\mciteSetBstMidEndSepPunct{\mcitedefaultmidpunct}
{\mcitedefaultendpunct}{\mcitedefaultseppunct}\relax
\EndOfBibitem
\bibitem[Wang \latin{et~al.}(2021)Wang, Liigand, Tian, Arndt, Greiner, and Wishart]{wang2021cfm}
Wang,~F.; Liigand,~J.; Tian,~S.; Arndt,~D.; Greiner,~R.; Wishart,~D.~S. CFM-ID 4.0: more accurate ESI-MS/MS spectral prediction and compound identification. \emph{Analytical chemistry} \textbf{2021}, \emph{93}, 11692--11700\relax
\mciteBstWouldAddEndPuncttrue
\mciteSetBstMidEndSepPunct{\mcitedefaultmidpunct}
{\mcitedefaultendpunct}{\mcitedefaultseppunct}\relax
\EndOfBibitem
\bibitem[Young \latin{et~al.}(2024)Young, Wang, Wishart, Wang, R{\"o}st, and Greiner]{young2024fragnnet}
Young,~A.; Wang,~F.; Wishart,~D.; Wang,~B.; R{\"o}st,~H.; Greiner,~R. FraGNNet: a deep probabilistic model for mass spectrum prediction. \emph{arXiv preprint arXiv:2404.02360} \textbf{2024}, \relax
\mciteBstWouldAddEndPunctfalse
\mciteSetBstMidEndSepPunct{\mcitedefaultmidpunct}
{}{\mcitedefaultseppunct}\relax
\EndOfBibitem
\bibitem[Goldman \latin{et~al.}(2024)Goldman, Li, and Coley]{goldman2024generating}
Goldman,~S.; Li,~J.; Coley,~C.~W. Generating molecular fragmentation graphs with autoregressive neural networks. \emph{Analytical Chemistry} \textbf{2024}, \emph{96}, 3419--3428\relax
\mciteBstWouldAddEndPuncttrue
\mciteSetBstMidEndSepPunct{\mcitedefaultmidpunct}
{\mcitedefaultendpunct}{\mcitedefaultseppunct}\relax
\EndOfBibitem
\bibitem[Nowatzky \latin{et~al.}(2025)Nowatzky, Russo, Lisec, Kister, Reinert, Muth, and Benner]{nowatzky2025fiora}
Nowatzky,~Y.; Russo,~F.~F.; Lisec,~J.; Kister,~A.; Reinert,~K.; Muth,~T.; Benner,~P. FIORA: Local neighborhood-based prediction of compound mass spectra from single fragmentation events. \emph{Nature Communications} \textbf{2025}, \emph{16}, 2298\relax
\mciteBstWouldAddEndPuncttrue
\mciteSetBstMidEndSepPunct{\mcitedefaultmidpunct}
{\mcitedefaultendpunct}{\mcitedefaultseppunct}\relax
\EndOfBibitem
\bibitem[Kojima \latin{et~al.}(2022)Kojima, Gu, Reid, Matsuo, and Iwasawa]{kojima2022large}
Kojima,~T.; Gu,~S.~S.; Reid,~M.; Matsuo,~Y.; Iwasawa,~Y. Large language models are zero-shot reasoners. \emph{Advances in neural information processing systems} \textbf{2022}, \emph{35}, 22199--22213\relax
\mciteBstWouldAddEndPuncttrue
\mciteSetBstMidEndSepPunct{\mcitedefaultmidpunct}
{\mcitedefaultendpunct}{\mcitedefaultseppunct}\relax
\EndOfBibitem
\bibitem[Wei \latin{et~al.}(2022)Wei, Wang, Schuurmans, Bosma, Xia, Chi, Le, Zhou, \latin{et~al.} others]{wei2022chain}
Wei,~J.; Wang,~X.; Schuurmans,~D.; Bosma,~M.; Xia,~F.; Chi,~E.; Le,~Q.~V.; Zhou,~D.; others Chain-of-thought prompting elicits reasoning in large language models. \emph{Advances in neural information processing systems} \textbf{2022}, \emph{35}, 24824--24837\relax
\mciteBstWouldAddEndPuncttrue
\mciteSetBstMidEndSepPunct{\mcitedefaultmidpunct}
{\mcitedefaultendpunct}{\mcitedefaultseppunct}\relax
\EndOfBibitem
\bibitem[Bran \latin{et~al.}(2023)Bran, Cox, Schilter, Baldassari, White, and Schwaller]{bran2023chemcrow}
Bran,~A.~M.; Cox,~S.; Schilter,~O.; Baldassari,~C.; White,~A.~D.; Schwaller,~P. Chemcrow: Augmenting large-language models with chemistry tools. \emph{arXiv preprint arXiv:2304.05376} \textbf{2023}, \relax
\mciteBstWouldAddEndPunctfalse
\mciteSetBstMidEndSepPunct{\mcitedefaultmidpunct}
{}{\mcitedefaultseppunct}\relax
\EndOfBibitem
\bibitem[Touvron \latin{et~al.}(2023)Touvron, Lavril, Izacard, Martinet, Lachaux, Lacroix, Rozi{\`e}re, Goyal, Hambro, Azhar, \latin{et~al.} others]{touvron2023llama}
Touvron,~H.; Lavril,~T.; Izacard,~G.; Martinet,~X.; Lachaux,~M.-A.; Lacroix,~T.; Rozi{\`e}re,~B.; Goyal,~N.; Hambro,~E.; Azhar,~F.; others Llama: Open and efficient foundation language models. \emph{arXiv preprint arXiv:2302.13971} \textbf{2023}, \relax
\mciteBstWouldAddEndPunctfalse
\mciteSetBstMidEndSepPunct{\mcitedefaultmidpunct}
{}{\mcitedefaultseppunct}\relax
\EndOfBibitem
\bibitem[Achiam \latin{et~al.}(2023)Achiam, Adler, Agarwal, Ahmad, Akkaya, Aleman, Almeida, Altenschmidt, Altman, Anadkat, \latin{et~al.} others]{achiam2023gpt}
Achiam,~J.; Adler,~S.; Agarwal,~S.; Ahmad,~L.; Akkaya,~I.; Aleman,~F.~L.; Almeida,~D.; Altenschmidt,~J.; Altman,~S.; Anadkat,~S.; others Gpt-4 technical report. \emph{arXiv preprint arXiv:2303.08774} \textbf{2023}, \relax
\mciteBstWouldAddEndPunctfalse
\mciteSetBstMidEndSepPunct{\mcitedefaultmidpunct}
{}{\mcitedefaultseppunct}\relax
\EndOfBibitem
\bibitem[Guo \latin{et~al.}(2024)Guo, Nan, Zhou, Guo, Guo, Surve, Liang, Chawla, Wiest, and Zhang]{guo2024can}
Guo,~K.; Nan,~B.; Zhou,~Y.; Guo,~T.; Guo,~Z.; Surve,~M.; Liang,~Z.; Chawla,~N.; Wiest,~O.; Zhang,~X. Can llms solve molecule puzzles? a multimodal benchmark for molecular structure elucidation. \emph{Advances in Neural Information Processing Systems} \textbf{2024}, \emph{37}, 134721--134746\relax
\mciteBstWouldAddEndPuncttrue
\mciteSetBstMidEndSepPunct{\mcitedefaultmidpunct}
{\mcitedefaultendpunct}{\mcitedefaultseppunct}\relax
\EndOfBibitem
\bibitem[Bhattacharya \latin{et~al.}(2024)Bhattacharya, Cassady, Hickner, and Reinhart]{bhattacharya2024large}
Bhattacharya,~D.; Cassady,~H.~J.; Hickner,~M.~A.; Reinhart,~W.~F. Large language models as molecular design engines. \emph{Journal of Chemical Information and Modeling} \textbf{2024}, \emph{64}, 7086--7096\relax
\mciteBstWouldAddEndPuncttrue
\mciteSetBstMidEndSepPunct{\mcitedefaultmidpunct}
{\mcitedefaultendpunct}{\mcitedefaultseppunct}\relax
\EndOfBibitem
\bibitem[Pei \latin{et~al.}(2023)Pei, Zhang, Zhu, Wu, Gao, Wu, Xia, and Yan]{pei2023biot5}
Pei,~Q.; Zhang,~W.; Zhu,~J.; Wu,~K.; Gao,~K.; Wu,~L.; Xia,~Y.; Yan,~R. Biot5: Enriching cross-modal integration in biology with chemical knowledge and natural language associations. \emph{arXiv preprint arXiv:2310.07276} \textbf{2023}, \relax
\mciteBstWouldAddEndPunctfalse
\mciteSetBstMidEndSepPunct{\mcitedefaultmidpunct}
{}{\mcitedefaultseppunct}\relax
\EndOfBibitem
\bibitem[Liu \latin{et~al.}(2023)Liu, Li, Luo, Fei, Cao, Kawaguchi, Wang, and Chua]{liu2023molca}
Liu,~Z.; Li,~S.; Luo,~Y.; Fei,~H.; Cao,~Y.; Kawaguchi,~K.; Wang,~X.; Chua,~T.-S. Molca: Molecular graph-language modeling with cross-modal projector and uni-modal adapter. \emph{arXiv preprint arXiv:2310.12798} \textbf{2023}, \relax
\mciteBstWouldAddEndPunctfalse
\mciteSetBstMidEndSepPunct{\mcitedefaultmidpunct}
{}{\mcitedefaultseppunct}\relax
\EndOfBibitem
\bibitem[Zhao \latin{et~al.}(2025)Zhao, Ma, Chen, Sun, Li, Xia, Chen, Xu, Zhu, Zhu, \latin{et~al.} others]{zhao2025developing}
Zhao,~Z.; Ma,~D.; Chen,~L.; Sun,~L.; Li,~Z.; Xia,~Y.; Chen,~B.; Xu,~H.; Zhu,~Z.; Zhu,~S.; others Developing ChemDFM as a large language foundation model for chemistry. \emph{Cell Reports Physical Science} \textbf{2025}, \emph{6}\relax
\mciteBstWouldAddEndPuncttrue
\mciteSetBstMidEndSepPunct{\mcitedefaultmidpunct}
{\mcitedefaultendpunct}{\mcitedefaultseppunct}\relax
\EndOfBibitem
\bibitem[Liu \latin{et~al.}(2024)Liu, Ding, Zhou, Fan, and Tan]{liu2024moleculargpt}
Liu,~Y.; Ding,~S.; Zhou,~S.; Fan,~W.; Tan,~Q. Moleculargpt: Open large language model (llm) for few-shot molecular property prediction. \emph{arXiv preprint arXiv:2406.12950} \textbf{2024}, \relax
\mciteBstWouldAddEndPunctfalse
\mciteSetBstMidEndSepPunct{\mcitedefaultmidpunct}
{}{\mcitedefaultseppunct}\relax
\EndOfBibitem
\bibitem[Chithrananda \latin{et~al.}(2020)Chithrananda, Grand, and Ramsundar]{chithrananda2020chemberta}
Chithrananda,~S.; Grand,~G.; Ramsundar,~B. ChemBERTa: large-scale self-supervised pretraining for molecular property prediction. \emph{arXiv preprint arXiv:2010.09885} \textbf{2020}, \relax
\mciteBstWouldAddEndPunctfalse
\mciteSetBstMidEndSepPunct{\mcitedefaultmidpunct}
{}{\mcitedefaultseppunct}\relax
\EndOfBibitem
\bibitem[Li \latin{et~al.}(2023)Li, Gao, Song, Wang, Xu, and Han]{li2023druggpt}
Li,~Y.; Gao,~C.; Song,~X.; Wang,~X.; Xu,~Y.; Han,~S. Druggpt: A gpt-based strategy for designing potential ligands targeting specific proteins. \emph{bioRxiv} \textbf{2023}, 2023--06\relax
\mciteBstWouldAddEndPuncttrue
\mciteSetBstMidEndSepPunct{\mcitedefaultmidpunct}
{\mcitedefaultendpunct}{\mcitedefaultseppunct}\relax
\EndOfBibitem
\bibitem[Liu \latin{et~al.}(2024)Liu, Guo, Li, Liu, Huang, Ke, and Lv]{liu2024drugllm}
Liu,~X.; Guo,~Y.; Li,~H.; Liu,~J.; Huang,~S.; Ke,~B.; Lv,~J. Drugllm: Open large language model for few-shot molecule generation. \emph{arXiv preprint arXiv:2405.06690} \textbf{2024}, \relax
\mciteBstWouldAddEndPunctfalse
\mciteSetBstMidEndSepPunct{\mcitedefaultmidpunct}
{}{\mcitedefaultseppunct}\relax
\EndOfBibitem
\bibitem[Shen \latin{et~al.}(2024)Shen, Zhou, and Che]{shen2024fragllama}
Shen,~J.; Zhou,~S.; Che,~X. FragLlama: Next-fragment prediction for molecular design. \emph{bioRxiv} \textbf{2024}, 2024--09\relax
\mciteBstWouldAddEndPuncttrue
\mciteSetBstMidEndSepPunct{\mcitedefaultmidpunct}
{\mcitedefaultendpunct}{\mcitedefaultseppunct}\relax
\EndOfBibitem
\bibitem[Su \latin{et~al.}(2025)Su, Chen, Jiang, Zhong, Wang, and Liu]{su2025language}
Su,~Y.; Chen,~J.; Jiang,~Z.; Zhong,~Z.; Wang,~L.; Liu,~Q. Language models can understand spectra: A multimodal model for molecular structure elucidation. \emph{arXiv preprint arXiv:2508.08441} \textbf{2025}, \relax
\mciteBstWouldAddEndPunctfalse
\mciteSetBstMidEndSepPunct{\mcitedefaultmidpunct}
{}{\mcitedefaultseppunct}\relax
\EndOfBibitem
\bibitem[Elhenawy \latin{et~al.}(2024)Elhenawy, Abdelhay, Alhadidi, Ashqar, Jaradat, Jaber, Glaser, and Rakotonirainy]{elhenawy2024eyeballing}
Elhenawy,~M.; Abdelhay,~A.; Alhadidi,~T.~I.; Ashqar,~H.~I.; Jaradat,~S.; Jaber,~A.; Glaser,~S.; Rakotonirainy,~A. Eyeballing combinatorial problems: A case study of using multimodal large language models to solve traveling salesman problems. International Conference on Intelligent Systems, Blockchain, and Communication Technologies. 2024; pp 341--355\relax
\mciteBstWouldAddEndPuncttrue
\mciteSetBstMidEndSepPunct{\mcitedefaultmidpunct}
{\mcitedefaultendpunct}{\mcitedefaultseppunct}\relax
\EndOfBibitem
\bibitem[Madaan \latin{et~al.}(2023)Madaan, Tandon, Gupta, Hallinan, Gao, Wiegreffe, Alon, Dziri, Prabhumoye, Yang, \latin{et~al.} others]{madaan2023self}
Madaan,~A.; Tandon,~N.; Gupta,~P.; Hallinan,~S.; Gao,~L.; Wiegreffe,~S.; Alon,~U.; Dziri,~N.; Prabhumoye,~S.; Yang,~Y.; others Self-refine: Iterative refinement with self-feedback. \emph{Advances in Neural Information Processing Systems} \textbf{2023}, \emph{36}, 46534--46594\relax
\mciteBstWouldAddEndPuncttrue
\mciteSetBstMidEndSepPunct{\mcitedefaultmidpunct}
{\mcitedefaultendpunct}{\mcitedefaultseppunct}\relax
\EndOfBibitem
\bibitem[Iklassov \latin{et~al.}(2024)Iklassov, Du, Akimov, and Takac]{iklassov2024self}
Iklassov,~Z.; Du,~Y.; Akimov,~F.; Takac,~M. Self-guiding exploration for combinatorial problems. \emph{Advances in Neural Information Processing Systems} \textbf{2024}, \emph{37}, 130569--130601\relax
\mciteBstWouldAddEndPuncttrue
\mciteSetBstMidEndSepPunct{\mcitedefaultmidpunct}
{\mcitedefaultendpunct}{\mcitedefaultseppunct}\relax
\EndOfBibitem
\bibitem[Bushuiev \latin{et~al.}(2024)Bushuiev, Bushuiev, de~Jonge, Young, Kretschmer, Samusevich, Heirman, Wang, Zhang, D{\"u}hrkop, \latin{et~al.} others]{bushuiev2024massspecgym}
Bushuiev,~R.; Bushuiev,~A.; de~Jonge,~N.~F.; Young,~A.; Kretschmer,~F.; Samusevich,~R.; Heirman,~J.; Wang,~F.; Zhang,~L.; D{\"u}hrkop,~K.; others MassSpecGym: A benchmark for the discovery and identification of molecules. \emph{arXiv preprint arXiv:2410.23326} \textbf{2024}, \relax
\mciteBstWouldAddEndPunctfalse
\mciteSetBstMidEndSepPunct{\mcitedefaultmidpunct}
{}{\mcitedefaultseppunct}\relax
\EndOfBibitem
\bibitem[Djeffal(2025)]{djeffal2025reflexive}
Djeffal,~C. Reflexive Prompt Engineering: A Framework for Responsible Prompt Engineering and AI Interaction Design. Proceedings of the 2025 ACM Conference on Fairness, Accountability, and Transparency. 2025; pp 1757--1768\relax
\mciteBstWouldAddEndPuncttrue
\mciteSetBstMidEndSepPunct{\mcitedefaultmidpunct}
{\mcitedefaultendpunct}{\mcitedefaultseppunct}\relax
\EndOfBibitem
\bibitem[Schulhoff \latin{et~al.}(2024)Schulhoff, Ilie, Balepur, Kahadze, Liu, Si, Li, Gupta, Han, Schulhoff, \latin{et~al.} others]{schulhoff2024prompt}
Schulhoff,~S.; Ilie,~M.; Balepur,~N.; Kahadze,~K.; Liu,~A.; Si,~C.; Li,~Y.; Gupta,~A.; Han,~H.; Schulhoff,~S.; others The prompt report: a systematic survey of prompt engineering techniques. \emph{arXiv preprint arXiv:2406.06608} \textbf{2024}, \relax
\mciteBstWouldAddEndPunctfalse
\mciteSetBstMidEndSepPunct{\mcitedefaultmidpunct}
{}{\mcitedefaultseppunct}\relax
\EndOfBibitem
\bibitem[Shen \latin{et~al.}(2024)Shen, Tenenholtz, Hall, Alvarez-Melis, and Fusi]{shen2024tag}
Shen,~J.; Tenenholtz,~N.; Hall,~J.~B.; Alvarez-Melis,~D.; Fusi,~N. Tag-LLM: Repurposing general-purpose LLMs for specialized domains. \emph{arXiv preprint arXiv:2402.05140} \textbf{2024}, \relax
\mciteBstWouldAddEndPunctfalse
\mciteSetBstMidEndSepPunct{\mcitedefaultmidpunct}
{}{\mcitedefaultseppunct}\relax
\EndOfBibitem
\bibitem[Shengyu \latin{et~al.}(2023)Shengyu, Linfeng, Xiaoya, Sen, Xiaofei, Shuhe, Jiwei, Hu, Tianwei, Wu, \latin{et~al.} others]{shengyu2023instruction}
Shengyu,~Z.; Linfeng,~D.; Xiaoya,~L.; Sen,~Z.; Xiaofei,~S.; Shuhe,~W.; Jiwei,~L.; Hu,~R.; Tianwei,~Z.; Wu,~F.; others Instruction tuning for large language models: A survey. \emph{arXiv preprint arXiv:2308.10792} \textbf{2023}, \relax
\mciteBstWouldAddEndPunctfalse
\mciteSetBstMidEndSepPunct{\mcitedefaultmidpunct}
{}{\mcitedefaultseppunct}\relax
\EndOfBibitem
\bibitem[Peeperkorn \latin{et~al.}(2024)Peeperkorn, Kouwenhoven, Brown, and Jordanous]{peeperkorn2024temperature}
Peeperkorn,~M.; Kouwenhoven,~T.; Brown,~D.; Jordanous,~A. Is temperature the creativity parameter of large language models? \emph{arXiv preprint arXiv:2405.00492} \textbf{2024}, \relax
\mciteBstWouldAddEndPunctfalse
\mciteSetBstMidEndSepPunct{\mcitedefaultmidpunct}
{}{\mcitedefaultseppunct}\relax
\EndOfBibitem
\bibitem[Krenn \latin{et~al.}(2020)Krenn, H{\"a}se, Nigam, Friederich, and Aspuru-Guzik]{krenn2020self}
Krenn,~M.; H{\"a}se,~F.; Nigam,~A.; Friederich,~P.; Aspuru-Guzik,~A. Self-referencing embedded strings (SELFIES): A 100\% robust molecular string representation. \emph{Machine Learning: Science and Technology} \textbf{2020}, \emph{1}, 045024\relax
\mciteBstWouldAddEndPuncttrue
\mciteSetBstMidEndSepPunct{\mcitedefaultmidpunct}
{\mcitedefaultendpunct}{\mcitedefaultseppunct}\relax
\EndOfBibitem
\bibitem[Krenn \latin{et~al.}(2022)Krenn, Ai, Barthel, Carson, Frei, Frey, Friederich, Gaudin, Gayle, Jablonka, \latin{et~al.} others]{krenn2022selfies}
Krenn,~M.; Ai,~Q.; Barthel,~S.; Carson,~N.; Frei,~A.; Frey,~N.~C.; Friederich,~P.; Gaudin,~T.; Gayle,~A.~A.; Jablonka,~K.~M.; others SELFIES and the future of molecular string representations. \emph{Patterns} \textbf{2022}, \emph{3}\relax
\mciteBstWouldAddEndPuncttrue
\mciteSetBstMidEndSepPunct{\mcitedefaultmidpunct}
{\mcitedefaultendpunct}{\mcitedefaultseppunct}\relax
\EndOfBibitem
\bibitem[Weininger(1988)]{weininger1988smiles}
Weininger,~D. SMILES, a chemical language and information system. 1. Introduction to methodology and encoding rules. \emph{Journal of chemical information and computer sciences} \textbf{1988}, \emph{28}, 31--36\relax
\mciteBstWouldAddEndPuncttrue
\mciteSetBstMidEndSepPunct{\mcitedefaultmidpunct}
{\mcitedefaultendpunct}{\mcitedefaultseppunct}\relax
\EndOfBibitem
\bibitem[Zheng \latin{et~al.}(2024)Zheng, Zhang, Zhang, Ye, Luo, Feng, and Ma]{zheng2024llamafactory}
Zheng,~Y.; Zhang,~R.; Zhang,~J.; Ye,~Y.; Luo,~Z.; Feng,~Z.; Ma,~Y. LlamaFactory: Unified Efficient Fine-Tuning of 100+ Language Models. Proceedings of the 62nd Annual Meeting of the Association for Computational Linguistics (Volume 3: System Demonstrations). Bangkok, Thailand, 2024\relax
\mciteBstWouldAddEndPuncttrue
\mciteSetBstMidEndSepPunct{\mcitedefaultmidpunct}
{\mcitedefaultendpunct}{\mcitedefaultseppunct}\relax
\EndOfBibitem
\bibitem[Hu \latin{et~al.}(2022)Hu, Shen, Wallis, Allen-Zhu, Li, Wang, Wang, Chen, \latin{et~al.} others]{hu2022lora}
Hu,~E.~J.; Shen,~Y.; Wallis,~P.; Allen-Zhu,~Z.; Li,~Y.; Wang,~S.; Wang,~L.; Chen,~W.; others Lora: Low-rank adaptation of large language models. \emph{ICLR} \textbf{2022}, \emph{1}, 3\relax
\mciteBstWouldAddEndPuncttrue
\mciteSetBstMidEndSepPunct{\mcitedefaultmidpunct}
{\mcitedefaultendpunct}{\mcitedefaultseppunct}\relax
\EndOfBibitem
\bibitem[Kim and Lee(2024)Kim, and Lee]{kim2024strategic}
Kim,~J.; Lee,~J. Strategic data ordering: Enhancing large language model performance through curriculum learning. \emph{arXiv preprint arXiv:2405.07490} \textbf{2024}, \relax
\mciteBstWouldAddEndPunctfalse
\mciteSetBstMidEndSepPunct{\mcitedefaultmidpunct}
{}{\mcitedefaultseppunct}\relax
\EndOfBibitem
\bibitem[Li \latin{et~al.}(2022)Li, Zhang, and He]{li2022stability}
Li,~C.; Zhang,~M.; He,~Y. The stability-efficiency dilemma: Investigating sequence length warmup for training GPT models. \emph{Advances in Neural Information Processing Systems} \textbf{2022}, \emph{35}, 26736--26750\relax
\mciteBstWouldAddEndPuncttrue
\mciteSetBstMidEndSepPunct{\mcitedefaultmidpunct}
{\mcitedefaultendpunct}{\mcitedefaultseppunct}\relax
\EndOfBibitem
\bibitem[Grattafiori \latin{et~al.}(2024)Grattafiori, Dubey, Jauhri, Pandey, Kadian, Al-Dahle, Letman, Mathur, Schelten, Vaughan, \latin{et~al.} others]{grattafiori2024llama}
Grattafiori,~A.; Dubey,~A.; Jauhri,~A.; Pandey,~A.; Kadian,~A.; Al-Dahle,~A.; Letman,~A.; Mathur,~A.; Schelten,~A.; Vaughan,~A.; others The llama 3 herd of models. \emph{arXiv preprint arXiv:2407.21783} \textbf{2024}, \relax
\mciteBstWouldAddEndPunctfalse
\mciteSetBstMidEndSepPunct{\mcitedefaultmidpunct}
{}{\mcitedefaultseppunct}\relax
\EndOfBibitem
\bibitem[Ridder \latin{et~al.}(2014)Ridder, van~der Hooft, and Verhoeven]{ridder2014automatic}
Ridder,~L.; van~der Hooft,~J.~J.; Verhoeven,~S. Automatic compound annotation from mass spectrometry data using MAGMa. \emph{Mass Spectrometry} \textbf{2014}, \emph{3}, S0033--S0033\relax
\mciteBstWouldAddEndPuncttrue
\mciteSetBstMidEndSepPunct{\mcitedefaultmidpunct}
{\mcitedefaultendpunct}{\mcitedefaultseppunct}\relax
\EndOfBibitem
\bibitem[Lo \latin{et~al.}(2023)Lo, Pollice, Nigam, White, Krenn, and Aspuru-Guzik]{lo2023recent}
Lo,~A.; Pollice,~R.; Nigam,~A.; White,~A.~D.; Krenn,~M.; Aspuru-Guzik,~A. Recent advances in the self-referencing embedded strings (SELFIES) library. \emph{Digital Discovery} \textbf{2023}, \emph{2}, 897--908\relax
\mciteBstWouldAddEndPuncttrue
\mciteSetBstMidEndSepPunct{\mcitedefaultmidpunct}
{\mcitedefaultendpunct}{\mcitedefaultseppunct}\relax
\EndOfBibitem
\bibitem[Landrum()]{landrum2013rdkit}
Landrum,~G. RDKit: open-source cheminformatics. \url{http://www.rdkit.org}, [Accessed on 11/1/2024]\relax
\mciteBstWouldAddEndPuncttrue
\mciteSetBstMidEndSepPunct{\mcitedefaultmidpunct}
{\mcitedefaultendpunct}{\mcitedefaultseppunct}\relax
\EndOfBibitem
\bibitem[Mandal \latin{et~al.}(2025)Mandal, Sasnauskas, and Radanovic]{mandal2025distributionally}
Mandal,~D.; Sasnauskas,~P.; Radanovic,~G. Distributionally robust reinforcement learning with human feedback. \emph{arXiv preprint arXiv:2503.00539} \textbf{2025}, \relax
\mciteBstWouldAddEndPunctfalse
\mciteSetBstMidEndSepPunct{\mcitedefaultmidpunct}
{}{\mcitedefaultseppunct}\relax
\EndOfBibitem
\bibitem[Kaufmann \latin{et~al.}(2024)Kaufmann, Weng, Bengs, and H{\"u}llermeier]{kaufmann2024survey}
Kaufmann,~T.; Weng,~P.; Bengs,~V.; H{\"u}llermeier,~E. A survey of reinforcement learning from human feedback. \emph{arXiv preprint arXiv:2312.14925} \textbf{2024}, \relax
\mciteBstWouldAddEndPunctfalse
\mciteSetBstMidEndSepPunct{\mcitedefaultmidpunct}
{}{\mcitedefaultseppunct}\relax
\EndOfBibitem
\bibitem[Wu \latin{et~al.}(2025)Wu, Zhou, Ziheng, Peng, Ye, Hu, Zhu, Qi, Yang, and Yang]{wu2025generalization}
Wu,~Y.; Zhou,~Y.; Ziheng,~Z.; Peng,~Y.; Ye,~X.; Hu,~X.; Zhu,~W.; Qi,~L.; Yang,~M.-H.; Yang,~X. On the generalization of sft: A reinforcement learning perspective with reward rectification. \emph{arXiv preprint arXiv:2508.05629} \textbf{2025}, \relax
\mciteBstWouldAddEndPunctfalse
\mciteSetBstMidEndSepPunct{\mcitedefaultmidpunct}
{}{\mcitedefaultseppunct}\relax
\EndOfBibitem
\bibitem[Gao \latin{et~al.}(2025)Gao, Zhu, Sui, Kong, Aldogom, Huang, Noori, Shamji, Parvataneni, Tsiligkaridis, \latin{et~al.} others]{gao2025democratizing}
Gao,~S.; Zhu,~R.; Sui,~P.; Kong,~Z.; Aldogom,~S.; Huang,~Y.; Noori,~A.; Shamji,~R.; Parvataneni,~K.; Tsiligkaridis,~T.; others Democratizing AI scientists using ToolUniverse. \emph{arXiv preprint arXiv:2509.23426} \textbf{2025}, \relax
\mciteBstWouldAddEndPunctfalse
\mciteSetBstMidEndSepPunct{\mcitedefaultmidpunct}
{}{\mcitedefaultseppunct}\relax
\EndOfBibitem
\bibitem[Ghosal \latin{et~al.}(2022)Ghosal, Majumder, Mihalcea, and Poria]{ghosal2022two}
Ghosal,~D.; Majumder,~N.; Mihalcea,~R.; Poria,~S. Two is better than many? binary classification as an effective approach to multi-choice question answering. \emph{arXiv preprint arXiv:2210.16495} \textbf{2022}, \relax
\mciteBstWouldAddEndPunctfalse
\mciteSetBstMidEndSepPunct{\mcitedefaultmidpunct}
{}{\mcitedefaultseppunct}\relax
\EndOfBibitem
\end{mcitethebibliography}
